# Low-loss superconducting resonant circuits using vacuum-gap-based microwave components

Katarina Cicak[1], Dale Li[1], Joshua A. Strong[1,2], Michael S. Allman[1,2], Fabio Altomare[1], Adam J. Sirois[1,2], Jed D. Whittaker[1,2], John D. Teufel[1], and Raymond W. Simmonds[1]

[1]*National Institute of Standards and Technology, Boulder, Colorado 80305, USA*
[2]*University of Colorado, Boulder, Colorado 80309, USA*

We have produced high-quality complex microwave circuits, such as multiplexed resonators and superconducting phase qubits, using a "vacuum-gap" technology that eliminates lossy dielectric materials. We have improved our design and fabrication strategy beyond our earlier work, leading to increased yield, enabling the realization of these complex circuits. We incorporate both novel vacuum-gap wiring crossovers (VGX) for gradiometric inductors and vacuum-gap capacitors (VGC) on chip to produce resonant circuits that have large internal quality factors (30,000<$Q_i$<165,000) at 50 mK, outperforming most dielectric-filled devices. Resonators with VGCs as large as 180 pF confirm single mode behavior of our lumped-element components.

Low-loss microwave circuit components are crucial for emerging superconducting quantum-based technologies. Resonators built from passive circuits are key for microwave nano-mechanics experiments[1], kinetic inductance[2] and transition edge[3] photon detector arrays, quantum-limited microwave amplifiers[4,5], superconducting quantum bits (qubits)[6], and quantum memories[7]. Researchers developing quantum-limited detectors and amplifiers, as well as qubits, are concerned about intrinsic noise sources and decoherence. Both are thought to be associated with unwanted two-level system (TLS) defects found in amorphous dielectric materials.[8-10]

The TLSs are believed to be microscopic in origin resulting from structural imperfections.[8] They disrupt the operation of quantum circuits via electric (and possibly magnetic) dipole coupling to electromagnetic fields. For superconducting phase qubits, this has led to short energy relaxation times[10], reduced measurement fidelity[11], and unwanted qubit interactions[12]. Measurements involving lumped-element $LC$ resonators using dielectric-filled parallel-plate capacitors have confirmed that TLSs absorb energy at low temperatures and low voltages.[10,13] Furthermore, TLSs within amorphous native oxides found on the surfaces of coplanar waveguide (CPW) resonators have also contributed to noise in superconducting kinetic inductance detectors.[9] Removing the TLS defects is a major challenge for advancing the performance of most superconducting circuits operating at low temperature in the quantum regime.

Motivated by the need for lossless cryogenic microwave components on chip, various strategies have been pursued to eliminate TLS defects. One strategy is to search for amorphous materials with a more highly constrained lattice with fewer TLSs. Simply increasing the silicon content in silicon-nitride dielectrics improves performance.[14] A survey of different materials[13] shows that achieving low-voltage, low-temperature dielectric loss tangents near $10^{-5}$ is possible with amorphous hydrogenated silicon. Another effective strategy is to use planarized components that rely on vacuum and a crystalline substrate dielectric. When fabricated on sapphire or undoped silicon, interdigitated capacitors (IDCs) and CPWs show good performance.[13,15] However, these distributed structures have large footprints, often spanning millimeters, and can suffer from non-ideal characteristics such as parasitic inductances and additional resonant modes.

In this work, we developed methods for fabricating lumped-element circuit components that confine electromagnetic fields mostly to vacuum, eliminating lossy amorphous dielectrics. We microfabricated superconducting parallel-plate capacitors with a small vacuum gap (VGC) providing a relatively small on-chip footprint with nearly ideal capacitor behavior. We combined VGCs with microfabricated superconducting gradiometric inductors with vacuum-gap wiring crossovers (VGX) to form lumped-element resonators. Resonant transmission measurements were used to characterize microwave losses and to demonstrate frequency-domain multiplexing with multiple resonators attached to a single CPW feedline. In addition, we successfully fabricated and operated phase qubits incorporating VGCs and VGXs.

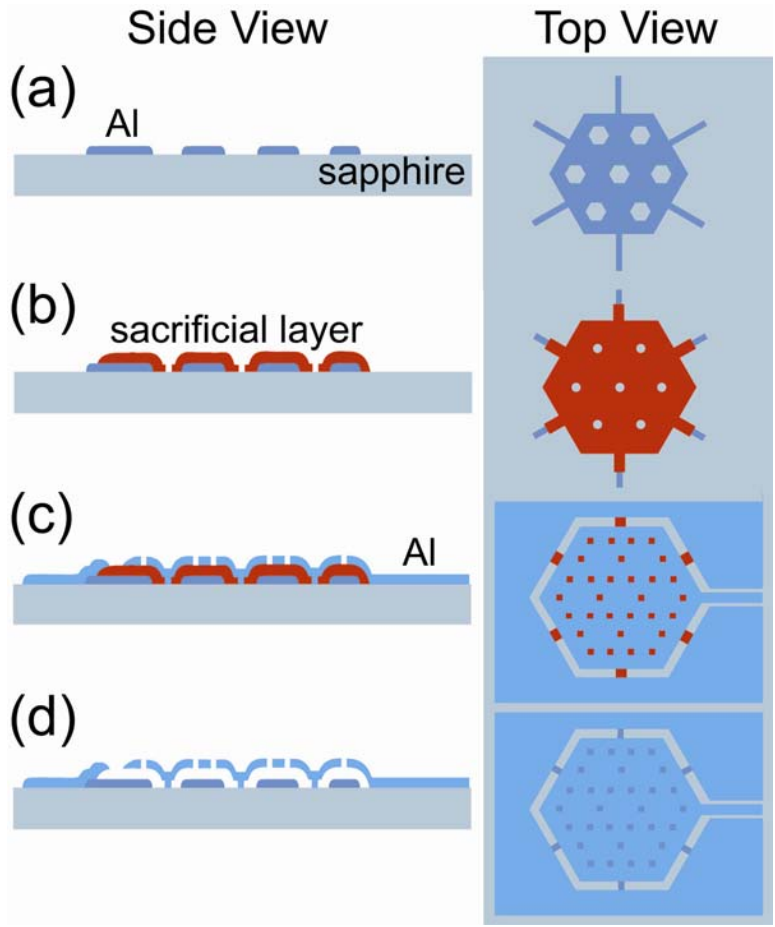

**FIG. 1.** (Color online) Fabrication procedure for VGCs. (a) Lower VGC plate patterned from an Al layer 100 nm thick on a sapphire wafer. (b) Patterned sacrificial layer (Nb, Si, or SiNx) 200 nm thick with "via" holes. (c) Al top capacitor plate 100 nm thick is sputtered and patterned on top of the sacrificial layer. Al fills the "via" holes in the sacrificial layer, forming posts that support the top plate. The lower plate is connected to the ground plane. (d) The sacrificial layer is removed by an isotropic high-pressure $SF_6$ plasma etch through the holes in the top plate.

Microfabrication of the VGC (shown in Fig. 1) and VGX structures is achieved with standard optical lithography and wet and dry thin-film etching techniques. A polycrystalline Al film, 100 nm thick, is sputtered on a sapphire wafer, and patterned by wet etching to form the lower capacitor plate perforated with a hexagonal hole pattern. Next, a 200 nm sacrificial layer is sputter- (Nb or Si) or plasma- ($SiN_x$) deposited and then patterned with "via" holes (2 μm diameter) aligned over the holes in the lower plate. A second layer of Al 100 nm thick is sputtered and patterned to form the top capacitor plate and the ground plane. An overlap contact connects the ground plane to the lower capacitor plate. Aluminum fills in the "via" holes, forming posts that support the top plate after the sacrificial layer is removed in a later step. An array of small holes (1 μm in diameter and offset from the via-posts) in the top plate facilitates the removal of the sacrificial layer after the wafer has been diced into 6.5×6.5 $mm^2$ chips. The sacrificial layer is removed with an isotropic high-pressure $SF_6$ plasma etch, leaving a gap between the plates. This fabrication process lithographically defines the via-type Al posts (unlike the previously developed VGCs[16]) eliminating the possibility of under- or over-etching the posts. We have achieved a significantly higher yield of non-shorted and non-collapsed capacitors, allowing us to build more complex circuits. VGXs are fabricated during the same sequence; sacrificial material sandwiched between crossing wires is removed, suspending the top wire. Away from the crossover, overlap contacts connect the top wire to the bottom wiring. Figure 2(a) shows an SEM image of a fabricated *LC* resonator containing a VGC (Fig. 2(b)) in parallel with an inductor with VGXs (Fig. 2(c)). We have also successfully integrated VGCs and VGXs fabrication with phase-qubit fabrication (see Fig. 2(d)).

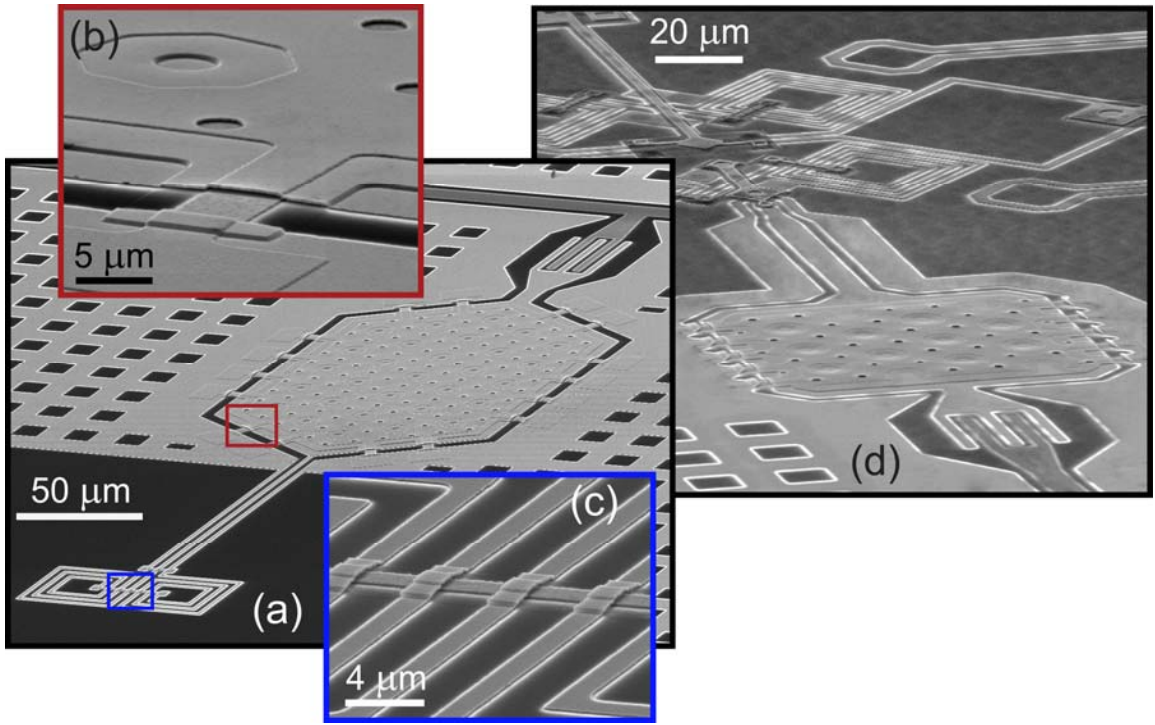

**FIG. 2.** (Color online) (a) SEM micrograph of a fabricated *LC* resonator incorporating a VGC and a gradiometric multicoil inductor with VGXs. The resonator is coupled to a CPW transmission line through an IDC and surrounded by a perforated ground plane. (b) Magnified part of the VGC where the gap between the top and bottom plates is visible. (c) Magnified part of the inductor showing VGXs. (d) Square VGC with Nb posts (previously developed) in parallel with a 6 $\mu m^2$ Al-AlO$_x$-Al Josephson junction as a part of a phase qubit.

Fabricated *LC* resonators coupled to a CPW transmission line with characteristic impedance $Z_0$ = 50 Ω through one or two small IDC coupling capacitors $C_C$<<$C$ (see Fig. 3(a,b)) are evaluated by microwave transmission measurements to obtain the quality factor of the resonator and to extract the intrinsic loss of the circuit components. The single-$C_C$ configuration allows us to frequency-division multiplex multiple resonators with different resonant frequencies. We used this configuration to characterize multiple sets of five resonators, with various values of *L*, *C*, and $C_C$, on a single chip with one microwave feedline per set, as shown in Fig. 3(c, d). All chips are wire-bonded to a shielded circuit board and cooled in vacuum to 50 mK. All resonators are driven by a synthesized signal generator through the input port 1. The transmitted power from the output port 2 is amplified at cryogenic temperatures (4 K) and then at room temperature before being measured with a spectrum analyzer (see Fig. 3(a, b)).

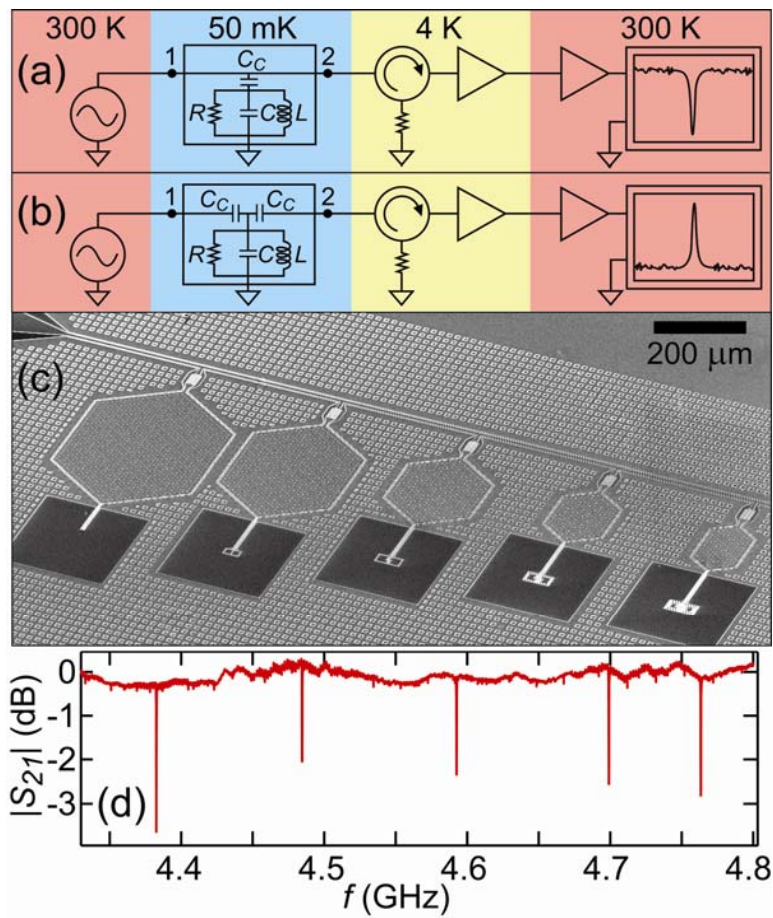

**FIG. 3.** (Color online) Measurement set-up to test *LC* resonators with (a) one or (b) two $C_C$. Microwave drive is attenuated by 20 dB at 4K and then by 20 dB at 50 mK before the input port 1 (not shown), which ensures a thermal population of much less than one photon in the resonator. (c) SEM micrograph of a fabricated circuit with five different *LC* resonators each connected through a $C_C$ to a single CPW transmission line. (d) Transmitted signal of a similar circuit with equal *C* design values, and *L* values staggered to produce a frequency separation between the resonance dips of ~100 MHz.

The impedance of a parallel *LC* resonator approaches infinity on resonance, and in the double-$C_C$ configuration the transmission spectrum exhibits a Lorentzian peak centered at frequency $f = f_R \approx (2\pi)^{-1}(LC)^{-1/2}$. The energy lost within the resonator can be modeled by an effective resistance *R* in parallel with *L* and *C*. In circuits with dielectric-filled parallel-plate capacitors, *R* is dominated by the loss in the capacitive component (see Fig. 4(c)). The loss tangent, $\tan(\delta)$, is related to the internal quality factor $Q_I$ of the resonator through $\tan(\delta) = 1/Q_I = (2\pi f_R R C)^{-1} \approx (L/C)^{1/2}/R$. $Q_I$ is related to the measured quality factor $Q_M$ and to $Q_C$, describing energy lost to the external feedlines, through $Q_M^{-1} = Q_I^{-1} + Q_C^{-1}$ with $Q_M = f_R/\delta f$, where $\delta f$ is the full width at half maximum of the resonant peak. For $C_C \ll C$, $Q_C \approx (C/C_C)/(4\pi f_R Z_0 C_C)$. A similar analysis of the single-$C_C$ circuit configuration, exhibiting a dip in the transmission spectrum, yields $Q_C \approx (C/C_C)/(\pi f_R Z_0 C_C)$.

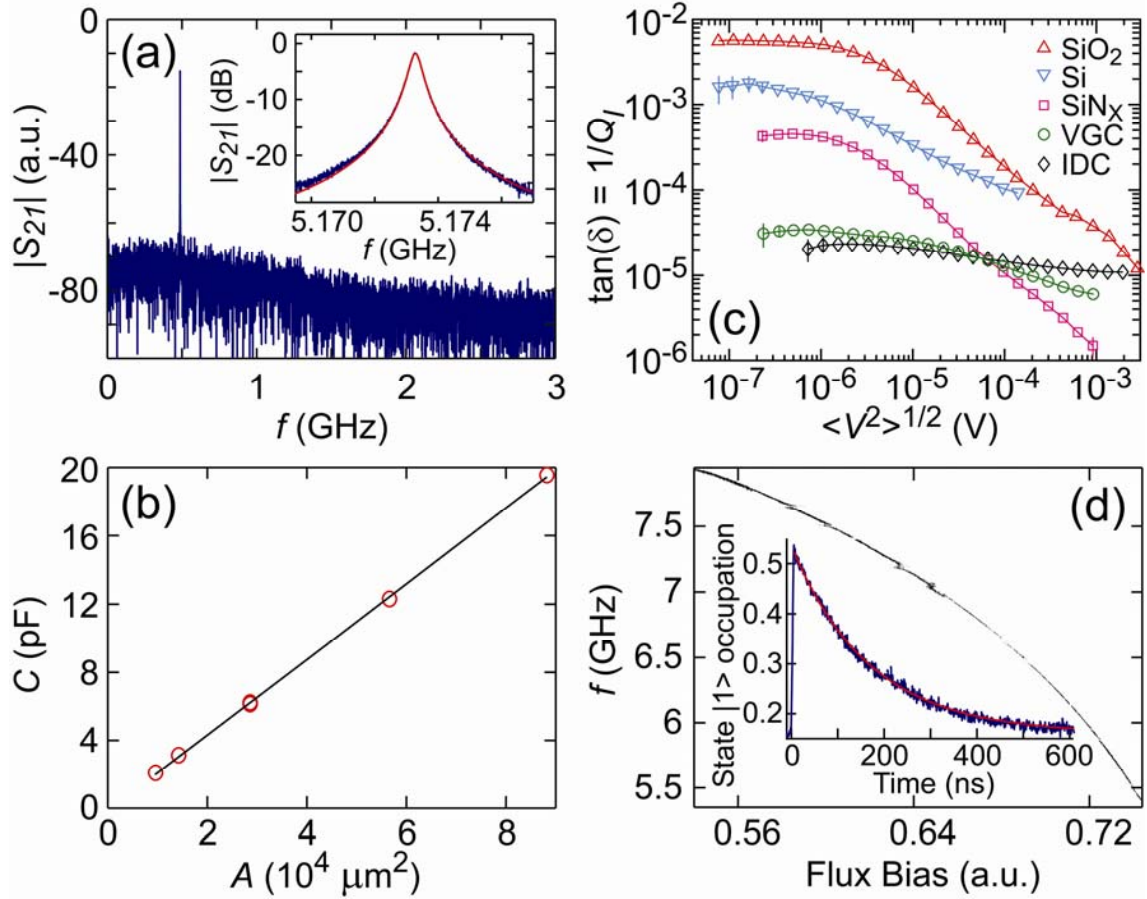

**FIG. 4.** (Color online) (a) Transmitted signal of an *LC* resonator with a large VGC (1 mm across and $C = 180$ pF). The data show a single resonance peak at $f_R \approx 500$ MHz and no higher-order resonant modes. The inset shows typical data (blue line) with a fitted Lorentzian peak (red line) for a double-$C_C$ circuit containing a VGC and VGXs. (b) Capacitance vs. VGC plate overlap area. (c) Measured tan($\delta$) vs. $\langle V^2 \rangle^{1/2}$ for three capacitors with dielectric materials ($SiO_2$, Si, or $SiN_x$) between the plates, a VGC with Al posts, and an IDC capacitor, all measured with double-$C_C$ circuits. (d) Spectroscopy of the phase qubit incorporating a VGC. The inset shows energy relaxation from the qubit excited state with characteristic time $T_1 \approx 170$ ns.

We have measured more than sixty resonant circuits with greater than 90 % yield. These structurally robust devices have survived multiple thermal cycles from room temperature to below 100 mK. They show single-mode resonances even for resonators constructed with large capacitors, 1 mm across and 180 pF (see Fig. 4(a)). Capacitance values *C* extracted from measured $f_R$ scale linearly with the total plate overlap area *A*, as expected (see Fig. 4(b)). However, we observe that the measured values of *C* are systematically larger than the design values $C_g = \varepsilon_0 A/d$, where $\varepsilon_0$ is the permittivity of free space and $d$ is the expected plate separation. Typically, $C/C_g \sim 4$. We believe that film stress during fabrication and thermal contractions during cooldown reduce $d$.[16] On a single chip, the spread in $C/C_g$ for eleven capacitors was less then 2 %, with a larger variation from chip to chip. Five nominally identical multiplexed resonators showed resonant frequencies near 4.5 GHz scattered over a bandwidth of less then 100 MHz. Thus with our high-$Q$ resonators and well defined $L$ and $C$ values, it is possible to narrowly space many resonances for multiplexing applications.

Loss tangent, $\tan(\delta)$, was extracted from the transmitted power spectra as a function of input power or the RMS voltage across the $LC$ resonator, $\langle V^2 \rangle^{1/2}$. Fig. 4(c) shows data from five resonators fabricated on sapphire substrates: three with common dielectric materials within parallel-plate capacitors, one with an IDC, and one with VGC and VGXs. The observed $\tan(\delta)$ increases at low powers, consistent with energy loss from TLSs.[10,13] Removing dielectric materials clearly improves device performance. IDC and VGC resonators show consistently weaker power dependence and lower overall $\tan(\delta)$ at the lowest powers (in the single-photon limit). In this range the VGC resonators had $\tan(\delta) \sim 3 \times 10^{-5}$, within a factor of 1.5 of the IDC device but with a footprint roughly ten times smaller on chip. Understanding what currently limits VGC resonator performance is still under investigation. We believe that amorphous native oxides on the inner surfaces of the VGC plates are a major contributor.[9] The parallel plate geometry of VGCs provides a good platform for further research, because the electric fields can be modeled easily. Fig. 4(d) shows a phase qubit spectrum for the device shown in Fig. 2(d) with an energy relaxation time of ~170 ns. The energy relaxation time of VGC-based qubits without the sacrificial layer removed is an order of magnitude worse. The dominating residual energy losses are associated with the Josephson junction[17], not the capacitor.

Using a "vacuum-gap" technology, we have produced high quality superconducting microwave resonators from novel lumped-element capacitive and inductive components. With small on-chip footprints and low losses these components make an attractive alternative to existing multiplexed resonators. They are well suited for many quantum-limited circuit applications including amplifiers, photon detectors, and qubits. We presented preliminary measurements for frequency-domain multiplexing of the $LC$ resonators and for the first time demonstrated the operation of phase qubits fabricated with VGCs and VGXs, the basic building blocks for the applications mentioned above. All devices show improved microwave characteristics at low temperatures, due to removal of dielectric materials. The parallel-plate construction of VGCs can provide a useful platform for future studies of loss mechanisms in microwave circuits.

This work was supported by NIST. Contribution of US government; not subject to copyright.